\documentclass{cupconf}
\usepackage{graphicx} 

  \checkfont{eurm10}
  \iffontfound
    \IfFileExists{upmath.sty}
      {\typeout{^^JFound AMS Euler Roman fonts on the system,
                   using the 'upmath' package.^^J}%
       \usepackage{upmath}}
      {\typeout{^^JFound AMS Euler Roman fonts on the system, but you
                   dont seem to have the}%
       \typeout{'upmath' package installed. cupconf.cls can take advantage
                 of these fonts,^^Jif you use 'upmath' package.^^J}%
      }
  \else
  \fi

  \checkfont{msam10}
  \iffontfound
    \IfFileExists{amssymb.sty}
      {\typeout{^^JFound AMS Symbol fonts on the system, using the
                'amssymb' package.^^J}%
       \usepackage{amssymb}%

      }{}
  \fi

  \IfFileExists{amsbsy.sty}
    {\typeout{^^JFound the 'amsbsy' package on the system, using it.^^J}%
     \usepackage{amsbsy}}
    {}

\newsavebox{\astrutbox}
\sbox{\astrutbox}{\rule[-5pt]{0pt}{20pt}}

\newcommand{\msun}{$M_{\odot}$}

\newcommand\etal{\mbox{et al.}}

\title[Stars and star clusters in the Antennae]{Massive stars and star clusters in the Antennae galaxies}

\author[B.~C.\ Whitmore]%
{B\ls R\ls A\ls D\ls L\ls E\ls Y\ns C.\ns W\ls H\ls I\ls T\ls M\ls O\ls R\ls E}

\affiliation{Space Telescope Science Institute, 3700 San Martin Drive, 
Baltimore, MD 21218, USA}

\begin{document}

\maketitle

\begin{abstract}

Large numbers of young stars are formed in merging galaxies, such as
the Antennae galaxies. Most of these stars are formed in compact
star clusters (i.e., super star clusters), which have been the focus
of a large number of  studies. However, an increasing number of
projects are beginning to
focus on the individual stars as well. In this contribution, we examine
a few results relevant to the triggering of star and star cluster formation;
ask what fraction of stars form in the field rather than in clusters;
and begin to explore the demographics of both the massive
stars and star clusters in the Antennae.

\end{abstract}

\firstsection 
\section{Introduction}

It is now well accepted that most star formation occurs in clustered
environments, such as associations, 
groups and clusters (e.g., Lada \& Lada 2003). In addition, it is
clear that star formation is greatly enhanced in merging galaxies,
making them an excellent place to study the formation of large
numbers of young, massive stars, albeit with the disadvantage of
having to work with stars at larger distances than the nearby groups and clusters
in our own galaxy. In keeping with their galactic counterparts, most of the
stars in merging galaxies also form in clusters, the brightest
and most compact of which have been dubbed ``super star clusters.'' 
Hence, understanding what triggers the formation of star clusters in 
mergers may be an important clue for understanding the formation of stars in general.

The excellent spatial resolution of the \emph{Hubble Space Telescope} 
(\emph{HST}) has rejuvenated the study of young star clusters in recent years 
(e.g., see reviews by Whitmore 2003, and Larsen 2005). One of the most
important results is that the brightest of the super star clusters have all the
attributes expected of young globular clusters (e.g., Holtzman 1992). An 
equally important result is that most of the groups and clusters do not 
appear to be bound, with roughly 90\% being dispersed into the field each 
decade of log time (i.e., ``infant mortality''; Whitmore 2003; Fall 2004; Fall, 
Chandar, \& Whitmore 2005).  Hence, understanding the destruction
of clusters may be the key to understanding the demographics of both
star clusters and field stars.

The Antennae galaxies (NGC~4038/39) are the nearest and youngest
prototypical merger in the Toomre (1977) sequence. Hence, they may be
our best chance for studying the formation of super star clusters and the 
massive stars within a major merger. While other galaxies will be briefly 
discussed at various parts of this review, the Antennae will be our 
centerpiece.  Figure~1 shows an example of some of the super star clusters 
in the Antennae (two left panels). Knot~S, shown in the upper left, will be 
the focus of several parts of this paper. The central cluster in Knot~S 
contains at least 10$^7$~\msun\/ alone, while the entire region contains 
well over 10$^8$~\msun.  Note that Knot~S consists of more than a 
single cluster. While it is difficult to distinguish individual supergiant stars 
in the outer region from faint clusters based on this image alone (this will 
be the main subject of \S4), at least a dozen objects are clearly resolved, 
and hence are sizeable clusters in their own right.  To provide some 
perspective, Figure~2 shows a superposition of what 30~Doradus 
(M$_V \approx -10$, $\approx$ 10$^5$~\msun) would look like at the 
distance of the Antennae.

\begin{figure}
\centering
\includegraphics[angle=0,width=\textwidth]{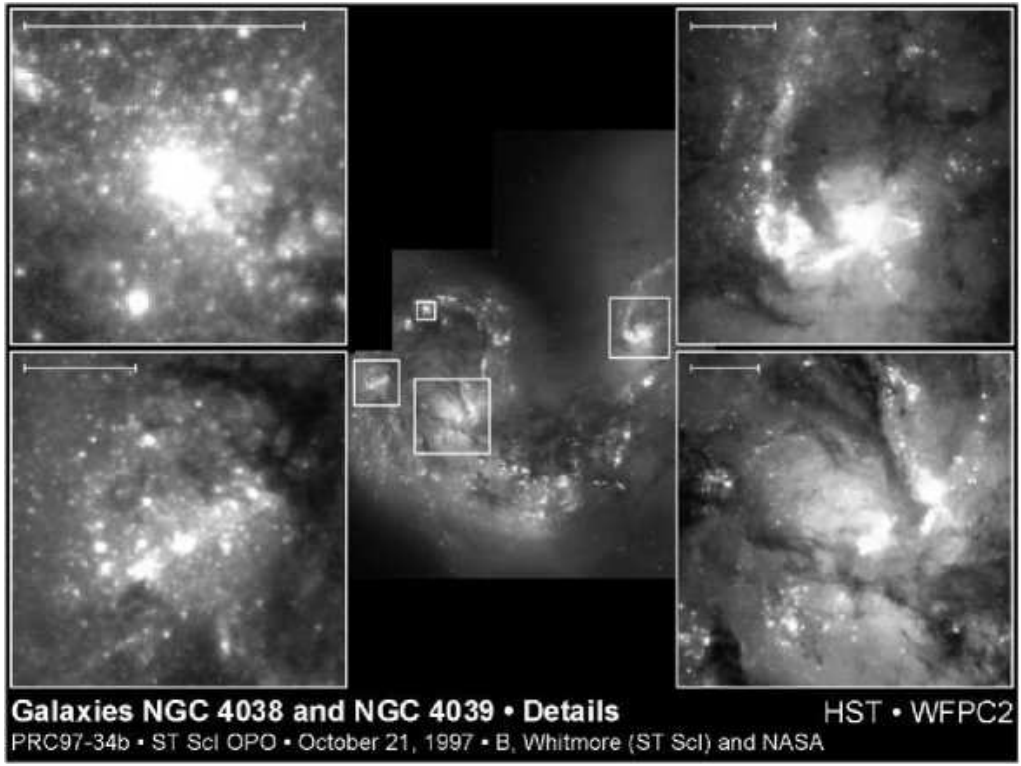}
\caption{Blowup of two of the brightest knots of clusters in the Antennae galaxies 
(left) and the central regions of the two galaxies (right) 
from Whitmore \etal\/ (1999).}
\label{fig1}
\end{figure}

While a great deal of attention has been paid to the study of super
star clusters in external galaxies during the past decade, relatively
little work has been done on the demographics of individual stars in these
galaxies. Reasons include the larger distance, which makes it difficult
to study anything but the brightest stars, and the high degree of
crowding due to the large number of stars and the clustered nature of
star formation. In general, it would seem that a detailed study of
nearby star formation regions, such as the Orion Nebula, would be more
fruitful. However, there are two basic reasons why it is important
to study individual stars in more distant galaxies as well. The first is the
opportunity to study larger samples of stars (e.g., $\approx10^7$ in 
Knot~S) in a specific cluster. This would allow us to determine whether  
the most massive star in a cluster is determined by statistics or physics 
(see Weidner \& Kroupa 2006; Elmegreen 2005; and Figer 2005 for 
discussions).  Another motivating factor is to determine whether there 
are two modes of star formation (i.e., violent and quiescent; Gallagher 
2004) which result in different stellar IMFs.

In this contribution we will first examine what has been learned about the 
triggering of star and star cluster formation in the Antennae. We will then 
address the question of whether essentially all stars form in clustered 
environments. We will also explore whether there is any evidence for 
an upper mass cutoff for the stellar IMF in the Antennae. Finally, we will 
describe an effort to develop a general framework for understanding the 
demographics of both stars and star clusters.

\begin{figure}
\centering
\includegraphics[angle=0,width=\textwidth]{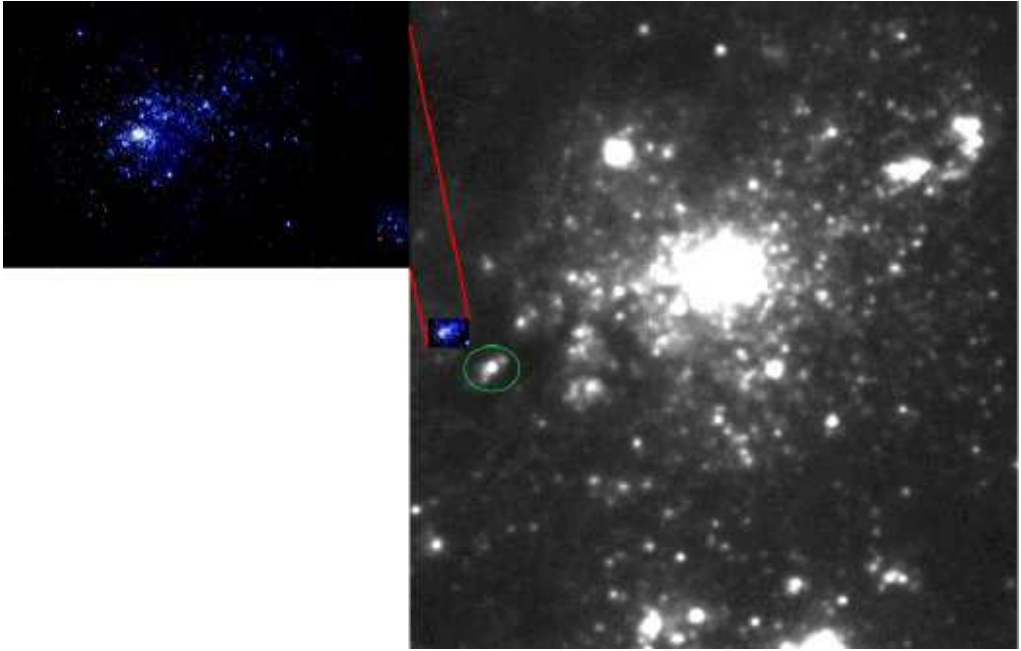}
\caption{What 30 Doradus (upper left) would look like at the distance of the
Antennae. While it would clearly be discernable as a cluster, it would be
dwarfed by Knot~S itself (the central object in the right-hand panel) and
would also be superseded by about a dozen other clusters in the region.}
\label{fig2}
\end{figure}

\section{What triggers the formation of star clusters (and hence stars) in
the Antennae?}

It is clear that shocks play an important role in triggering star formation. 
However, what is not clear is how they do this. One popular mechanism for 
triggering star formation in merging galaxies has been high-velocity 
cloud-cloud collisions (e.g., Kumai, Hashi, \& Fujimoto 1993 suggest that 
collisions with relative velocities in the range 50--100~km~s${-1}$ are required).

Whitmore \etal\/ 2005 obtained long-slit spectroscopy using STIS on 
\emph{HST} to address this question. They found that the velocity
fields are remarkably quiescent, with RMS dispersions less than about
10~km~s$^{-1}$, essentially the same as in the disks of normal spiral
galaxies (Figure~3). This does not support models that rely on 
high-velocity cloud-cloud collisions as the triggering mechanism, but is 
consistent with models where a high pressure interstellar medium 
implodes the GMCs without greatly affecting their initial velocity 
distribution (e.g., Jog \& Solomon 1992). This also supports earlier 
results (Zhang, Fall, \& Whitmore 2001) that found essentially no 
correlation between star cluster formation and the velocity gradients 
and dispersions of H$\alpha$, H~\textsc{i}, or CO.
In retrospect, this is also evident from the existence of a large
number of young clusters in the disk-like regions of NGC~4038, which
still has a relatively quiescent, disk-like rotation curve (see Amram 
\etal\/ 1992). 

\begin{figure}
\centering
\includegraphics[angle=-90,scale=0.7]{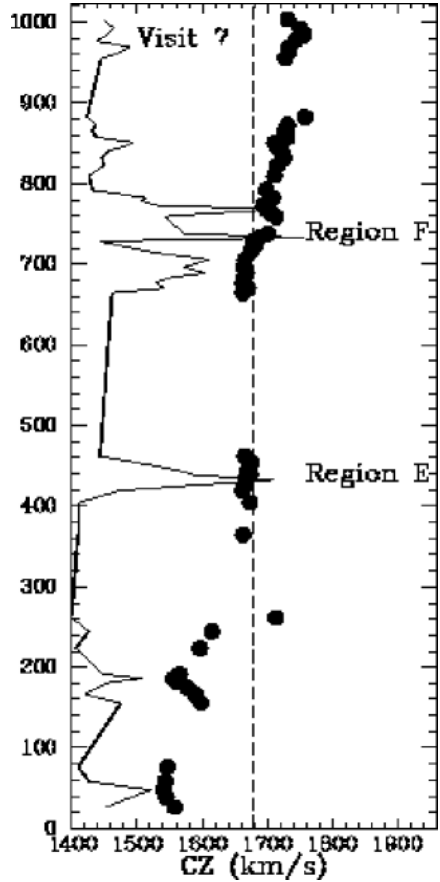}
\caption{H$\alpha$ velocities based on long-slit observations using 
the STIS detector on \emph{HST}. Note how small the velocity 
dispersion is for a given subregion (e.g., $\approx$10~km~s$^{-1}$ for 
region~F, once the large scale gradient is removed). See Whitmore 
\etal\/ 2005 for details.  }
\label{fig3}
\end{figure}

Another approach is to look for evidence of triggering by age-dating
clusters and looking for a pattern of older star formation (the initial
burst) surrounded by younger star formation (more recent bursts). Evidence 
for this has been seen around 30~Doradus (e.g., Walborn \etal\/ 1999), 
with the youngest star formation at the tips of ``pillars" pointing back towards
the central object. 

Perhaps the central question here is not whether star formation can 
be triggered by previous star formation, which it clearly can, but how 
important this effect is (Whitmore 2003). Put another way, is local 
triggering more important, or global triggering? Few attempts have 
been made to try to quantify this. Figure~4 shows some evidence for
sequential triggering around Knot~S of the Antennae, with a clump of 
older clusters near the center ($>$10~Myr, circles), intermediate-age 
clusters further out (3--10~Myr, crosses), and a few very young clusters 
still further out ($<$3~Myr, squares). We note that the clear clumping of 
the different ages shows that we are able to measure ages reasonably 
well, at least on a relative scale. If there was a very large amount of 
scatter in determining the ages we would find the different ages more 
randomly arranged in Figure~4. 

In general, the fraction of luminosity in succeeding generations of star 
clusters appears to continuously decrease (i.e., each new generation 
does not produce a comparable generation, so that the process 
cannot continue in equilibrium). Hence, we conclude that triggered
star formation is a significant, but not dominant, component of the
overall star formation in the Antennae. More global processes, such 
as interactions and spiral arms, appear to be the primary drivers.

\begin{figure}
\centering
\includegraphics[angle=0,scale=0.5]{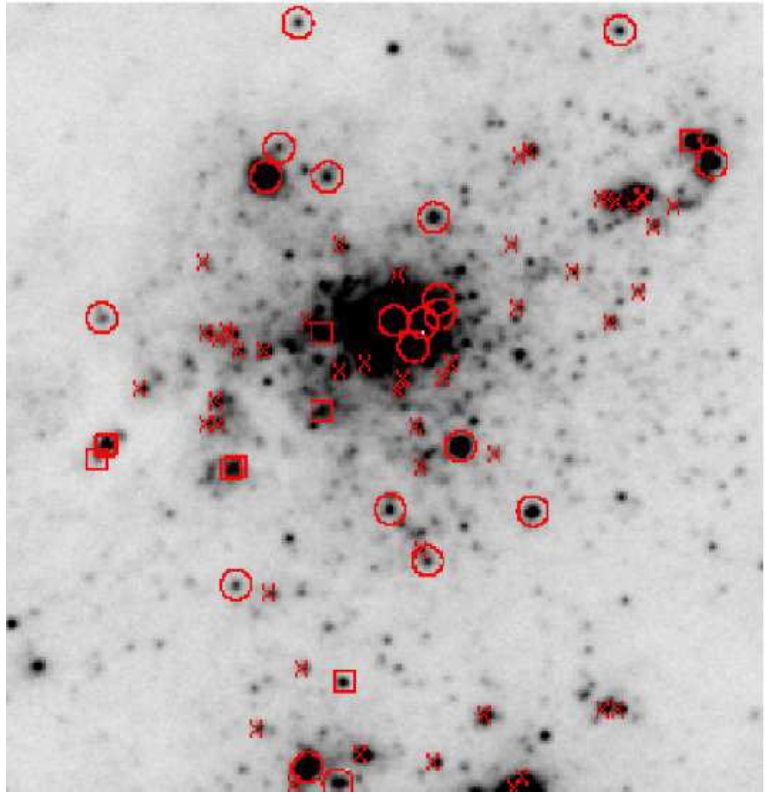}
\caption{Age estimates of clusters in Knot~S. Circles are for 
$>$10~Myr. Crosses are for 3--10~Myr. Squares are for $<$3~Myr. }
\label{fig6}
\end{figure}

\section{What fraction of stars are formed in clusters?}

It is well recognized that for the Milky Way, most stars are formed in
associations, groups and clusters (e.g., Lada \& Lada 2003).  De~Wit
et~al.\ (2005) provide a recent demonstration of this. They use proper
motions from \emph{Hipparcos} to estimate that only 4$\pm$2\% of the 
O~and B~stars in the Milky Way formed outside of groups or clusters (i.e.,
most of the O~and B~stars in the field are consistent with being
runaway stars from nearby groups.)

What have we learned on this subject from external galaxies, and in 
particular from the Antennae? Several early studies of star clusters in merger 
and starburst galaxies found that 10--50\% of the UV light (i.e., young stars) 
are found in clusters (e.g., Meuer 1995; Zepf \etal\/ 1999; Whitmore \& Zhang 
2002). The initial fraction of stars in clusters is even higher than these
estimates, since at least some clusters don't survive. In fact, as we shall see 
in \S5 (also see Fall, Chandar \& Whitmore 2005), we believe that roughly 
80--90\% of clusters disperse or are destroyed each decade of log time. 
Furthermore, our model, which incorporates this effect, predicts that if \emph{all 
stars are formed in clusters}, the amount of UV light we should observe in 
clusters should be $\approx$8\% for the Antennae, in good agreement with 
observations ($\approx$9\%; Whitmore \& Zhang 2002). See Fall, Chandar, \& 
Whitmore (2005) for a related calculation using total H$\alpha$ flux, again 
concluding that the observations are consistent with the idea that essentially 
all stars are formed in groups or clusters.

A related question is: What are the relative fractions of stars formed in 
associations, open clusters, and super star clusters? This is a difficult 
question to answer for a variety of reasons. First, there is no clear dividing 
line between these types of groupings, (i.e., they probably represent a 
continuum rather than distinct modes). Second, the objects are barely 
resolved, and are often found in very crowded regions, making it difficult 
to reliably separate the objects into more than a single bin. In addition, it 
is not clear how diffuse an open cluster or association needs to be before 
it falls out of the sample because it cannot be detected. It is interesting to 
note, however, that we seem to be able to account for essentially all of 
the UV light in the Antennae by stars that originally formed in groups and 
clusters (either still existing within clusters we detect, or from stars where 
the cluster has already dispersed). This suggests that a large fraction of 
stars are not formed in very diffuse associations that would be too faint to 
be in our sample.

We should also keep in mind that even clusters that survive will lose a 
large fraction of their stars from their outer halos.  For example, Whitmore 
\etal\/ 1999 found that young clusters like Knot~S have linear profiles, 
while older clusters have tidally truncated profiles, implying the removal of 
a large fraction of light from the outer regions (see Schweizer 2004 for a 
review on the sizes and radial profiles of clusters).  In fact, we estimate 
that $\approx$50\% of the light in Knot~S falls beyond 50~pc from the 
center, a typical tidal radius for a globular cluster.

Bastian \& Goodwin (2005) find similar profiles for the young clusters in 
M82, N1569, and N1705. They suggest that these profiles are compatible 
with $N$-body simulations of clusters with rapid removal of mass due to 
gas expulsion, hence supporting the basic interpretation that a large 
fraction of stars from clusters will eventually find themselves in the field. 
Fall, Chandar, \& Whitmore (2005) make a similar argument to explain 
the high infant mortality rate of clusters in the Antennae.

Comparisons between UV spectra from clusters, and from the diffuse 
field stars between clusters, provides another line of reasoning that 
supports this basic picture. For example, Chandar \etal\/ (2005) find 
that the integrated spectrum of the field stars in several local starburst 
galaxies is consistent with formation of the stars within clusters which 
dissolve with typical time scales of 7--10~Myr.  

\section{What can we learn about the stellar content of the super star 
clusters in the Antennae?}

In Whitmore \etal\/ (1999), one of our primary difficulties was 
differentiating stars from clusters. This led us to conclude that the 
number of young star clusters in the Antennae was between 800 and 
8000---a pretty big range! Our new ACS data, with its better spatial 
resolution, provides a better opportunity for making this determination 
and for studying the stars in their own right.

An important  tool we are employing in this analysis is a 
maximum-likelihood SED-fitting software package named
CHORIZOS, which is described in Ma\'{\i}z-Apell\'aniz (2004).  
Ubeda, Ma\'{\i}z-Apell\'aniz, \& MacKenty (2006) employed CHORIZOS 
to analyze \emph{HST} observations in six filter bands (F170, F336W, 
F555W, F814W, J, H) of NGC~4214, a nearby (3~Mpc) starburst dwarf 
galaxy.  Their main conclusions are: 1)~extinction is quite patchy, but 
relatively low around all but the youngest clusters, 2)~10 of the 12 clusters 
they studied have ages $<$10~Myr (note that this supports the infant 
mortality discussion that will be described in \S1 and \S5), 3)~the 
blue-to-red supergiant ratios are consistent with theory, 4)~the stellar IMF 
in the field is steeper than $-2.8$. This study is a good example of how
researchers are starting to study both the stellar and cluster contents of 
external galaxies. In the current paper, we use CHORIZOS to estimate 
values of M$_\mathrm{bol}$ and T$_\mathrm{e}$ for candidate stars in 
the Antennae.

\begin{figure}
\centering
\includegraphics[width=\textwidth]{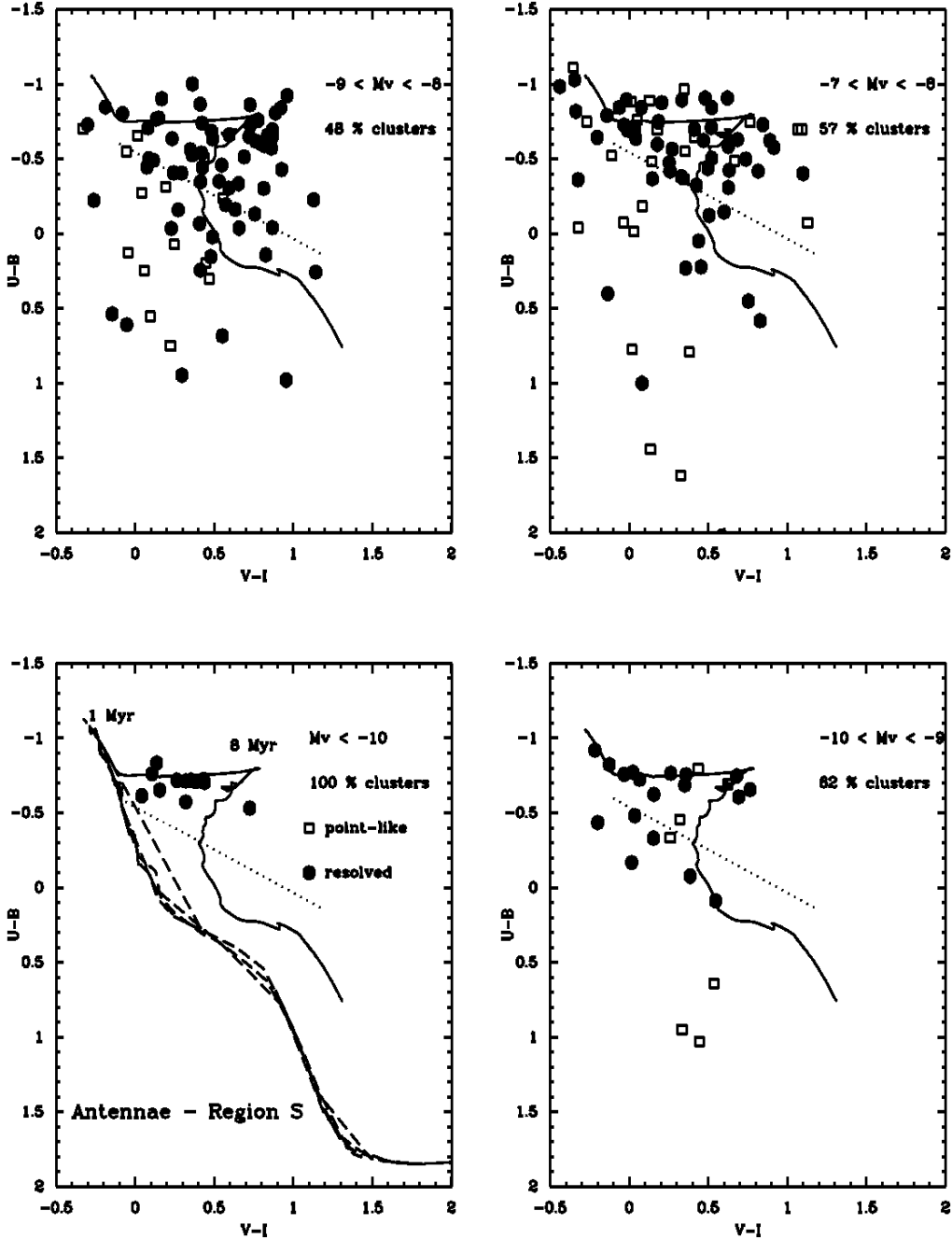}
\caption{$U-B$ vs.\ $V-I$ color-color diagram for four luminosity ranges
for a sample of objects around Knot~S (from Whitmore \etal\/ 2006; see
text for details). 
}
\label{fig5}
\end{figure}

We first ask the question: How well can we distinguish clusters from stars  
in Knot~S of the Antennae galaxies, based only on a concentration index 
(i.e., the luminosity of an object inside a 3~pixel radius compared to the 
luminosity inside a 1~pixel radius)? Figure~5 shows four luminosity ranges 
drawn from the sample of point-like objects around Knot~S, starting with 
the brightest objects  (M$_V < -10$;  bottom left),  and ranging to the fainter 
objects ($-7 < M_V < -8$; in the upper right). The objects with profiles 
indistinguishable from stars are shown as open squares, while the resolved 
objects are solid circles. The data is plotted on a $U-B$ vs.\ $V-I$ color-color 
diagram with Bruzual \& Charlot (2003) solar metallicity models superposed 
on all four panels using solid lines (young clusters are in the upper left and
old clusters in the lower  right; locations for 1- and 8-Myr clusters are shown 
on the bottom left figure). Padova models of stars brighter than M$_V=-7$ 
are shown by the dashed lines in the bottom left panel. The dotted line 
shows the reddening vector, and also acts as a rough dividing line between 
``cluster-space" (upper right) and ``star-space" (lower left). This works 
because essentially all of the objects in this region are young, hence there 
are no clusters that populate the bottom part of the Bruzual \& Charlot
cluster models. 

Several conclusions can be drawn from this figure. The first is that
if we select only the brightest objects (i.e., M$_V < -10$), they are all 
consistent with being young clusters ($<$8~Myr) with relatively little
extinction (i.e., they are very close to the Bruzual-Charlot models).
This is reassuring, since the brightest stars might be expected to be 
fainter than M$_V \approx -9$ (i.e., the brightest stars in the Milky Way; 
Humphreys 1983). This is the value we---and several other
researchers---have used to conservatively identify clusters in the past
(e.g., Whitmore \etal\/ 1999). 

If we cut the sample at M$_V< -9$ (lower right panel), things get a little 
more interesting. Near the bottom of the diagram we now have three 
point-like objects in Knot~S in the part of the diagram appropriate for 
stars. These all happen to have values of M$_V \approx -9.1$, just 
slightly brighter than our boundary condition between stars and clusters. 
We also find three point-like objects in or near cluster-space. This is 
our second important result, that while the concentration index is useful 
for telling the difference between clusters and stars, it is only partially 
successful. It appears that some clusters (based on their position in 
color-color space) are so concentrated that they cannot be 
distinguished from stars based on their size alone. This is also 
apparent from the fainter bins (upper panels), where a majority of the 
point-like objects are found in star-space, but a fair fraction are 
also found in cluster-space.

Another interesting point is that while most of the resolved objects
in the $-10 <$~M$_V< -9$ diagram hug the Bruzual-Charlot models
very nicely, about a half-dozen objects are just below the dotted line
used to separate cluster- and star-space. We believe most of these are
cases where there is a mixture of light from both a cluster and from
one or two bright stars in the cluster (i.e., if you added the light
from a cluster sitting on the Bruzual-Charlot cluster track and one of
the three stars at the bottom of the diagram, which have roughly the
same brightness, the result would be an object with an intermediate
color). This does not happen for the brightest clusters (i.e., with
M$_V < -10$) because these clusters have enough stars that one or
two random bright stars cannot greatly affect the total color.  Hence,
a certain degree of ``stochasticity" appears for young clusters with
magnitudes around M$_V \approx -9$ (i.e., masses around 
10$^4$~\msun). This effect has already been noted by other authors 
such as Cervino, Valls-Gabaud, \& Mass-Hesse (2002). Ubeda,
Ma\'{\i}z-Apell\'aniz, \& MacKenty (2006) also show a nice example 
of a cluster that appears to have both a blue and a red supergiant
superposed.

Our fourth, and perhaps most important result, is that roughly 50\%
of the objects fainter than M$_V < -9$ are clusters, based on their position
in the color-color diagram. This is actually a very conservative lower limit,
since, as we just noted, some of the objects around the dividing line are 
likely to be clusters with one or two stars pulling them just below the 
dividing line. This has a number of important ramifications, the most 
important being that it shows that the number of faint clusters continues 
to rise in a power law fashion. This provides a counter example to the 
claims that the initial cluster mass function in some galaxies may be a 
Gaussian (e.g., de~Grijs, Parmentier, \& Lamers 2005), since a Gaussian 
would require that essentially all the faint objects were individual stars.
Another ramification is that the quantity of clusters in the Antennae
numbers in the thousands, rather than the hundreds.

\begin{figure}
\centering
\includegraphics[angle=0,width=3.5in]{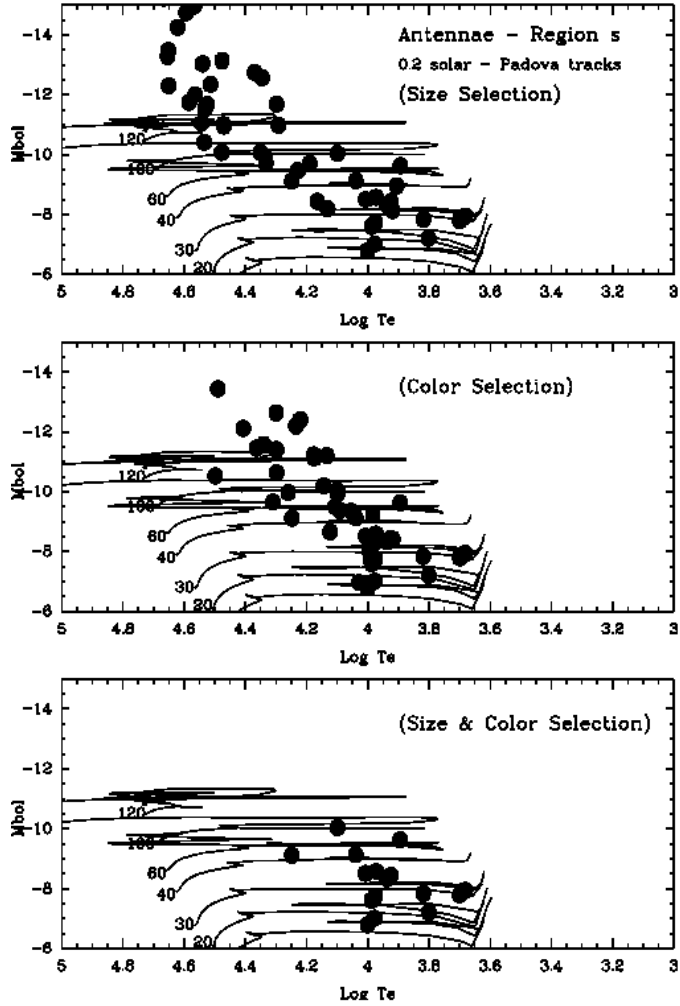}
\caption{M$_\mathrm{bol}$ vs.\ Log~T$_\mathrm{e}$ for candidate stars in Knot~S, 
using only a size selection, only a color selection, or both a size and color
selection. Note that using a combination of the size and color criteria 
does a good job of removing the clusters from the stars (i.e., the
remaining objects are in the part of the diagram expected for stars).
See text for details.}
\label{fig6}
\end{figure}

Hence, neither the concentration index (i.e., size), nor the position in the 
color-color diagram alone is completely successful in separating stars 
and clusters. What if we use a combination of both criteria? Figure~6 
(M$_\mathrm{bol}$  vs.\ Log~T$_\mathrm{e}$ diagrams for candidate massive 
stars around Knot~S) shows that this appears to work fairly well. Using 
\textit{either} the size or color criteria alone implies the existence of
stars that are more massive than the stellar tracks (two upper panels in 
Figure~6). However, if we use \textit{both} criteria simultaneously, all the 
remaining objects are consistent with being normal stars. We might note 
that this also suggests that there is an upper limit to the maximum mass 
of a star, since we would expect more massive stars in such a large 
sample of stars if the stellar IMF was a simple power law (see Weidner 
\& Kroupa 2006; Elmegreen 2005; and Figer 2005 for detailed 
treatments of this issue). This result should be considered tentative, 
however, pending a more careful look at other knots in the Antennae 
galaxies, and the development of Monte-Carlo simulations that will allow 
us to more quantitatively determine the statistical significance of the result. 

\section{The big picture---A general framework for understanding the
demographics of stars and star clusters}

The most extreme super star clusters, with magnitudes M$_V\approx -17$ 
and masses\break $\approx$$10^8$~\msun, are found in merging galaxies. One 
might therefore assume that there is something special about the physical 
environment in these galaxies that makes it possible to form such massive 
clusters there, but nowhere else. This suggests that there may be two 
modes of star cluster formation; one for relatively quiescent galaxies, such 
as normal spiral galaxies, and one for starbursting galaxies (e.g., Gallagher 
2004). However, the discovery of super star clusters in spiral galaxies by 
Larsen \& Richtler (1999)---and the subsequent demonstrations by Whitmore 
(2003; originally presented in 2000 as astro-ph/0012546), and Larsen 
(2002)---that there is a continuous correlation between the magnitude of the 
brightest cluster and the number of clusters (e.g., Figure~10 from Whitmore 
2003), suggests that there may be a single universal mode of star cluster 
formation, with the correlation simply being due to statistics. Hence, mergers 
and starburst galaxies may have the brightest clusters only because they 
have the most clusters. Several recent papers (e.g., Hunter \etal\/ 2003) 
have also realized that it is important to take this  ``size-of-sample'' effect 
into consideration when interpreting results.

\begin{figure}
\centering
\includegraphics[angle=0,scale=0.6]{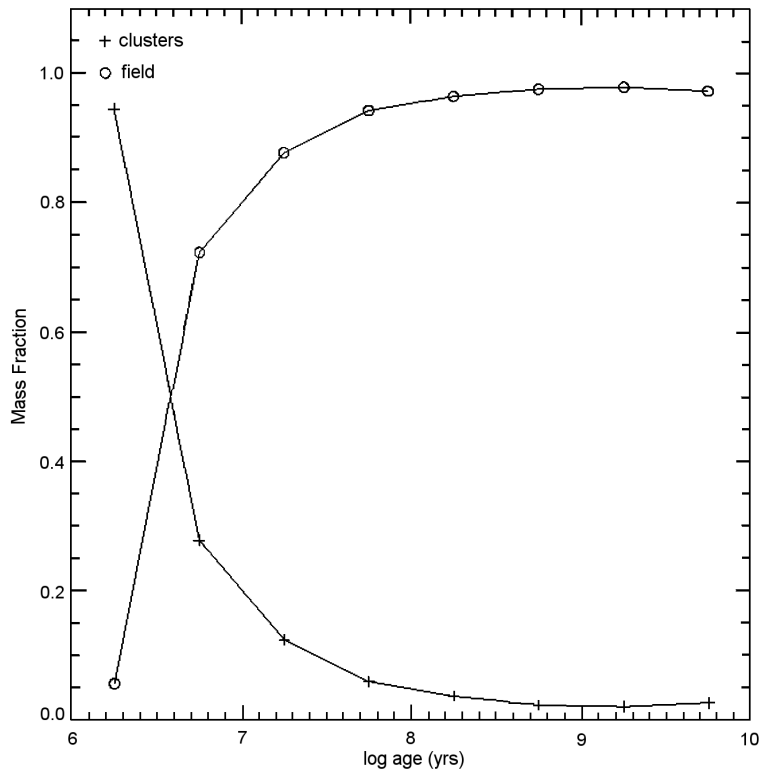}
\caption{Fraction of mass in clusters and in field stars as a function of time 
for our Antennae model. See text for details.}
\label{fig7}
\end{figure}
 
Similarly, there may be a universal power law relationship for the disruption  
rate of clusters. For example, Fall \etal\/ (2005) find that roughly 90\% of the 
clusters in the Antennae are removed from the sample each decade of log 
time (i.e., a power law with index  $-1$). Whitmore, Chandar, \& Fall (2006) 
show that this relationship appears to be the same for the Antennae, the 
SMC (data from Rafelski \& Zaritsky 2005), and the Milky Way (Lada \& 
Lada 2003).

These two results have motivated us to develop a general framework for 
understanding the demographics of both star clusters and the field stars, 
which we assume are formed as a by-product of the disrupted clusters 
(Whitmore 2004; Whitmore, Chandar, \& Fall 2006). The ingredients for the 
model are:
\begin{enumerate}
\item[1)]a universal initial mass function (power law, index $-2$);
\item[2)]various star (cluster) formation histories that can be coadded (e.g.,
constant, Gaussian, burst, ...);
\item[3)]various cluster disruption mechanisms 
     (e.g., $\tau^{-1}$  for $<$100~Myr, i.e., infant mortality; 
constant mass loss  for $>$100~Myr, i.e., 2-body relaxation); and
\item[4)]convolution with observational artifacts and selection effects.
\end{enumerate}

This simple model allows us to predict a wide variety of properties for the
clusters, field stars, and integrated properties of a galaxy. Of particular
relevance for the present paper is the agreement between prediction and
observations of what fraction of the UV light emitted by clusters (see discussion
in \S3). Figure~7 shows how the fraction of mass in clusters and in field
stars varies as a function of time for our best-fitting model of the Antennae.
We plan to extend this treatment to a number of other nearby galaxies including
M51, M101, and M82 (Chandar \& Whitmore 2006).

\begin{acknowledgements}
The author would like to thank several collaborators for their contributions
to a variety of projects that are mentioned in this review, in particular,
Rupali Chandar, Francois Schweizer, Mike Fall, Qing Zhang, and Barry Rothberg.
\end{acknowledgements}

\end{document}